\documentclass[11pt,superscriptaddress, onecolumn, nofootinbib]{revtex4-2}
\usepackage{graphicx}
\usepackage{amssymb,amsmath}
\usepackage{comment}
\usepackage{natbib}
\begin{document}
\title{Complex scalar field in $\kappa$-Minkowski spacetime
\thanks{Presented at POTOR 8}%
}

\author{Andrea Bevilacqua}
\email{andrea.bevilacqua@ncbj.gov.pl}
\affiliation{National Centre for Nuclear Research, Pasteura 7, 02-093
	Warszawa, Poland}


\begin{abstract}
	It is often expected that one cannot treat spacetime as a continuous manifold as the Planck scale is approached, because of to possible effects due to a quantum theory of gravity. There have been several proposals to model such a deviation from the classical behaviour, one of which is noncommutativity of spacetime coordinates. In this context, the non-commutativity scale is seen as an observer-independent length scale. Of course, such a scale impose a modification of ordinary relativistic symmetries, which now need to be deformed to accommodate this fundamental scale. The $\kappa$-Poincar\'e algebra is an example of this deformation.  In what follows I will briefly describe a construction of a $\kappa$-deformed complex scalar field theory, while at the same time shedding light on the behaviour of discrete and continuous symmetries in this formalism. This in turn will open the way to the study of the application of this formalism to actual physical processes. I will then conclude with some comments and prospects for the future.
\end{abstract}
  \maketitle
\section{From non-commutative spacetime to the action}

We will present a brief preview of the works in Refs.\ \cite{ref1}, \cite{ref2}. For a more detailed introduction to the framework and the formalism, see for example Ref.\ \cite{ref3}. The starting point is $k$-Minkowski spacetime, whose coordinates satisfy the $\mathfrak{an}(3)$ Lie algebra defined as $[\hat{x}^0, \hat{\mathbf{x}}^i] = i\hat{\mathbf{x}}^i/\kappa$. Notice that one recovers canonical Minkowski spacetime in the formal limit $\kappa\rightarrow\infty$. To have a more physical understanding of the above commutator, notice that $1/\kappa$ has dimensions of length, and it is sometimes identified with the Plank length. In order to build fields in this spacetime, we need to define plane waves first. To do so, one can proceed as follows. First pick an explicit representation of the $\mathfrak{an}(3)$ Lie algebra in terms of matrices (it turns out that the lowest dimensional one has dimension 5). Then, one can define a plane wave as an element $\hat{e}_k$ of the relative $AN(3)$ Lie group, i.e. for example\footnote{A different choice on the ordering results in a change of coordinates in momentum space.} $\hat{e}_k = \exp (i \mathbf{k}_i \hat{\mathbf{x}}^i) \exp (i k_0\hat{x}^0)$. Notice that, since the elements of the matrices $\hat{x}^\mu$ are dimensionful, so are the parameters $k_\mu$. Written explicitly, one has
\begin{equation}\label{II.1.19a}
	\hat  e_k   =\left(\begin{array}{ccc}
		\frac{ \bar p_4}\kappa \;&\;  \frac{\mathbf k}\kappa \; &\;
		\frac{  p_0}\kappa\\&&\\
		\frac{\mathbf p}\kappa  & \mathbf{1} & \frac{\mathbf p}\kappa  \\&&\\
		\frac{\bar p_0}\kappa\; & -\frac{\mathbf k}\kappa\; &
		\frac{  p_4}\kappa
	\end{array}\right)
\end{equation}
where\footnote{The explicit expression of $\bar{p}_4$ and $\bar{p}_0$ in terms of $k_\mu$ is not relevant for the present discussion.}
\begin{align}
	p_0
	=
	\kappa\sinh \frac{k_0}{\kappa}
	+
	\frac{\mathbf{k}^2}{2\kappa} e^{k_0/\kappa} 
	\quad
	\mathbf{p}_i
	=
	k_i e^{k_0/\kappa} 
	\quad
	p_4 
	= 
	\kappa\cosh \frac{k_0}{\kappa}
	-
	\frac{\mathbf{k}^2}{2\kappa} e^{k_0/\kappa}.
\end{align}
Both $k_\mu$ and $p_A(k)$ ($A=0,1,2,3,4$) are two different coordinates of momentum space. Notice that the $p_A$ are not independent, since $-p_0 + \mathbf{p}^2 + p_4^2 = \kappa^2$ (notice also that $p_+ = p_0+p_4 >0$). It is therefore clear that momentum space is curved. This is reflected in a deformed sum of momenta, which is defined through the group property $\hat{e}_k \hat{e}_l =: \hat{e}_{k\oplus l}$. At the same time, inverse momenta are defined in terms of group inverses $\hat{e}_k^{-1} =: \hat{e}_{S(k)}$, and $S(k)$ is called the antipode of $k$. In order to simplify the treatment of integrals and derivatives, we can use a\footnote{There are several equivalent ways in which a suitable Weyl map can be chosen, see Ref. \ \cite{ref1} for more details.} Weyl map $\mathcal{W}$ to send group elements $\hat{e}_k$ into canonical plane waves $e_p := \exp(ip_\mu (k) x^\mu)$. The group law is preserved thanks to the $\star$ product defined by $\mathcal{W}(\hat{e}_{k\oplus l}) = e_{p(k)\oplus q(l)} =: e_{p(k)}\star e_{q(l)}$. In general, the $\star$ product is not commutative. Because of this, we choose the following action for a $\kappa$-deformed complex scalar field.
\begin{equation}
	S = \frac{1}{2} \int_{\mathbb{R}^4} d^4x [(\partial^\mu\phi)^\dag \star (\partial_\mu\phi) + (\partial^\mu\phi) \star (\partial_\mu\phi)^\dag - m^2 (\phi^\dag \star \phi + \phi\star \phi^\dag)].
	\label{aba:1}
\end{equation}
To obtain the equations of motion one usually integrates by parts, but in this deformed context derivatives do not satisfy the Leibniz rule. In fact, if they did, one could get the following contradiction. 
\begin{align}
	i(p\oplus q)_\mu e_{p\oplus q}
	&=
	\partial_\mu (e_p \star e_q)
	=
	(\partial_\mu e_p) \star e_q
	+
	e_p \star \partial_\mu e_q =
	i(p+q)e_{p\oplus q}.
\end{align}
Hence one has to use the appropriate deformations for the Leibniz rules. After some computations, one can verify that the equations of motion satisfied by the field are the canonical Klein-Gordon equations, and the field which satisfies them can be written as
\begin{equation}
	\phi(x)=
	\int \frac{d^3p}{\sqrt{2\omega_p}} \xi(p) a_\mathbf{p}
	e^{-i(\omega_pt - \mathbf{p}\mathbf{x})}
	+
	\int \frac{d^3p^*}{\sqrt{2|\omega_p^*|}} \xi(p) b^\dag_{\mathbf{p}^*}
	e^{i(S(\omega_p)t - S(\mathbf{p})\mathbf{x})}.
	\label{aba:2}
\end{equation}
The action of the discrete transformations $C,P,T$ can be defined in the usual way in the deformed context, i.e. $T\phi(t,\mathbf{x})T^{-1}=\phi(-t,\mathbf{x})$, $P\phi(t,\mathbf{x})P^{-1}=\phi(t,-\mathbf{x})$, and 
$C\phi(t,\mathbf{x})C^{-1}=\phi^\dag(t,\mathbf{x})$. Notice that this is due to the presence of the antipode in the on-shell field. Furthermore, the action is manifestly invariant under both CPT and $\kappa$-deformed Lorentz transformations.

\section{Charges and features of the model}

We now need to compute the charges for our model. There are two main ways to do so. The first is by direct computation using the Noether theorem. However, the fact that derivatives do not follow the Leibniz rule makes this job a prohibitively difficult one, except for the case of translation charges. The second way is to use the covariant phase-space formalism. Although very direct, this second method needs to be carefully defined in
the deformed context. We chose a hybrid approach, namely we derived the translation charges from the Noether theorem, and then we built a covariant phase space approach which was able to reproduce the translation charges, allowing then to compute the remaining charges. Given a symplectic form $\Omega$, and a symmetry in spacetime described by the vector field $\xi$, the charge $Q_\xi$ can be defined by $-\delta_\xi \lrcorner \, \Omega = \delta Q_\xi$. In this context $\delta_\xi$ is a vector field in phase space describing the variation $\delta_\xi A$ of any physical quantity $A$ in phase space under the action of the symmetry $\xi$ in spacetime, $\delta$ is the exterior derivative in phase space, and $\lrcorner$ represents a contraction. The translation charges computed directly from the Noether theorem are the following\footnote{There is also a fifth charge $\mathcal{P}_4$, see \cite{ref1} for further details.}:
\begin{align}\label{directP}
	\mathcal{P}_\mu
	=
	\int d^3p \,
	\alpha(p)
	[-S(p)_\mu a_\mathbf{p}^\dag a_\mathbf{p} + p_\mu b_{\mathbf{p}^*}^\dag b_{\mathbf{p}^*}].
\end{align}
The quantity $\alpha(p)$ is a function of momenta whose explicit expression does not concern us in this context. To reproduce Eq. \eqref{directP} in the covariant phase space
formalism, we need to assume the following deformation of the canonical contraction rule between vector fields and forms\footnote{Recall that the canonical relation is $\delta_v \lrcorner \, (A\wedge B) = (\delta_v A) B - A(\delta_v B)$.}:
\begin{align}
	\delta_v \lrcorner \, (A\wedge B) 
	= (\delta_v A) B + A(S(\delta_v) B),
\end{align}
where $S(\delta_v)$ can be appropriately defined \cite{ref2}. Using this definition, one can compute all the remaining charges, and the creation/annihilation operators algebra (which turns out to be the canonical one). For example, the boost charge $\mathcal{N}_i$ is 
\begin{align}
	{\scriptstyle
	\mathcal{N}_i
	=
	- \frac{1}{2}
	\int d^3p \,\alpha\,
	\Bigg\{
	S(\omega_p) \left[
	\frac{\partial a_\mathbf{p}^\dag}{\partial S(\mathbf{p})^i}
	a_\mathbf{p}\,
	-
	a_\mathbf{p}^\dag \, \frac{\partial a_\mathbf{p}}{\partial S(\mathbf{p})^i}
	\right] +
	\omega_p
	\left[
	b_{\mathbf{p}}
	\frac{\partial b_{\mathbf{p}}^\dag}{\partial \mathbf{p}^i}
	-
	\frac{\partial b_{\mathbf{p}}}{\partial \mathbf{p}^i}
	b_{\mathbf{p}}^\dag
	\right]
	\Bigg\}
	}.
\end{align}
One can then verify that the charges satisfy the usual non-deformed Poincaré algebra. However, due to the effects of $\kappa$-deformation, one can also verify that, e.g, $[\mathcal{N}_i , C] \neq 0$, meaning that CPT symmetry is subtly violated. More explicitly, using the definition
\begin{align}
	C =
	\int \, d^3p
	\,
	(
	b^\dag_{\mathbf{p}}
	a_\mathbf{p}
	+
	a^\dag_{\mathbf{p}}
	b_{\mathbf{p}}
	),
\end{align}
one can show that $[\mathcal{N}_i , C]$ is given by
\begin{align}
	&
	\frac{i}{2}
	\int  d^3p 
	\Bigg\{
	S(\omega_p)
	\left[
	\frac{\partial a_{\mathbf{p}}}{\partial S(\mathbf{p})^i}
	b^\dag_{\mathbf{p}}
	-
	a_{\mathbf{p}}
	\frac{\partial b^\dag_{\mathbf{p}} }{\partial S(\mathbf{p})^i}
	+
	\frac{\partial a_{\mathbf{p}}^\dag}{\partial S(\mathbf{p})^i}
	b_{\mathbf{p}}
	-
	a_{\mathbf{p}}^\dag
	\frac{\partial b _{\mathbf{p}}}{\partial S(\mathbf{p})^i}
	\right] + \nonumber \\
	&\qquad \qquad +
	\omega_p
	\left[
	\frac{\partial b_{\mathbf{p}}^\dag}{\partial \mathbf{p}^i}
	a_\mathbf{p}
	-
	b_{\mathbf{p}}^\dag
	\frac{\partial a_\mathbf{p}}{\partial \mathbf{p}^i}
	+
	\frac{\partial b_{\mathbf{p}}}{\partial \mathbf{p}^i}
	a^\dag_{\mathbf{p}}
	-
	b_{\mathbf{p}}
	\frac{\partial a_\mathbf{p}^\dag}{\partial \mathbf{p}^i}
	\right]
	\Bigg\}.
\end{align}
Notice that in the limit $\kappa\rightarrow\infty$ one recovers the canonical result $[\mathcal{N}_i,C]=0$.

\section{Comments and conclusions}

The fact that $[\mathcal{N}_i,C]\neq 0$ has several important physical consequences. The most apparent one is a difference in the decay time between particles and antiparticles. Furthermore, we now have a well defined theory which will allow us to study in details the propagator and the $n$-point functions in general. From this point of view, it will be interesting to tackle the loops in this deformed context. Finally, what has been done for the case of the complex scalar field will be extended to fields of higher spins, in order to expand the discussion to more realistic phenomena.

\section*{Acknowledgments}
Parts of these works were supported by funds provided by
the Polish National Science Center, the project number 2019/33/B/ST2/00050.


\begin{thebibliography}{xx}
	
	
	\bibitem{ref1}
	M.\ Arzano, A.\ Bevilacqua, J.\ Kowalski-Glikman, G.\ Rosati, and J.\ Unger,
	Phys.\ Rev.\ D \ {\bf 103}, 106015 (2021)
	
	\bibitem{ref2}
	A.\ Bevilacqua, J.\ Kowalski-Glikman, and W.\ Wislicki,
	Phys.\ Rev.\ D \ {\bf 105}, 105004 (2022)
	
	\bibitem{ref3}
	M.\ Arzano, and J.\ Kowalski-Glikman,
	{\it Deformations of Spacetime Symmetries: Gravity, Group-Valued Momenta, and Non-Commutative Fields},
	Lecture Notes in Physics, Springer, 2021.
	
\end{thebibliography}
\end{document}